\documentclass[aps]{revtex4}

\usepackage{graphicx}
\usepackage{color}
\usepackage{topcapt}
\usepackage{booktabs}
\usepackage{amsmath, amsthm, amssymb}

\begin{document}

\bibliographystyle{apsrev}

\title{Semi-classical limit and minimum decoherence in the Conditional Probability Interpretation of Quantum Mechanics}
\author{Vincent Corbin and Neil J. Cornish}

\affiliation{Department of Physics, 
  Montana State University, Bozeman, MT 59717}

\begin{abstract}
The Conditional Probability Interpretation of Quantum Mechanics replaces the abstract notion of time used in standard Quantum Mechanics by
the time that can be read off from a physical clock. The use of physical clocks leads to apparent non-unitary and decoherence. Here we show
that a close approximation to standard Quantum Mechanics can be recovered from conditional Quantum Mechanics for semi-classical clocks,
and we use these clocks to compute the minimum decoherence predicted by the Conditional Probability Interpretation.
\end{abstract}

\pacs{}
\maketitle

\section{Introduction}

In Quantum Mechanics, each measurable quantity is associated with a quantum operator, and therefore is subject to quantum fluctuations
and to an uncertainty relation with its canonical conjugate. All except one, time. Time in Quantum Mechanics (and position in
Quantum Field Theory) has a special role. There is no time operator, no fluctuation, and the time-energy uncertainty is of a different
nature than the position-momentum uncertainty~\cite{Messiah}. In Quantum Mechanics, time is classical. One can measure time precisely
without affecting the system. One can measure time repeatedly without any consequences whatsoever. In the view of the quantization of
Gravity, where spacetime becomes a quantum dynamical variable, this is not acceptable. 
Many theories have been developed to try and fix this ``problem of time''. One in particular, the Conditional Probability Interpretation, is of
special interest to us. It was first developed by Page and Wootters~\cite{wooters}, and was recently refined by Dolby~\cite{dolby}.
There, time as we know it in Quantum Mechanics, becomes a parameter - some kind of internal time that one cannot measure. Instead one chooses
a quantum variable, which will be used as a ``clock''. Then the probability of measuring a certain value for a variable at time $t$, is
replaced by the probability of measuring this value when we have measured the ``clock'' variable at a given value. This interpretation is
quite natural in every day experiments. Indeed, one never directly measures time, but instead reads it through a clock. What we really
measure is the number of swings a pendulum makes, how many particles decay or other similar physical processes. Conventional Quantum Mechanics
would then only arise when taking the limit in which the ``clock'' behaves classically. 

The Conditional Probability Interpretation predicts effects absent from the standard Quantum theory. In particular, it predicts a non-unitarity
with respect to the variable chosen for ``time'', and the presence of decoherence in the system under study which leads to a loss of
information~\cite{relational_solution}. It turns out that using any physical clock will lead to this phenomenon. An estimate of the minimum decoherence has
been made by Gambini, Porto, and Pullin~\cite{fundamental_decoherence}, but, to the best of our knowledge, it has never been calculated directly
from the Conditional Probability Interpretation, without resorting to results from standard Quantum Mechanics. Our goal is to provide
a direct calculation of the minimum decoherence.

A brief background of the Conditional Probability Interpretation is given in Section 2. Section 3 describes in details how standard Quantum
Mechanics arises from the Conditional Probability Interpretation, in which limits and for what kind of clock. We also talk about the
semi-classical regime of a simple ``free particle clock'', which represents the simplest possible physical clock. In Section 4, we calculate
the minimum decoherence one can achieve using the ``free particle clock'', and compare our results with previous estimates. Conclusions are
presented in Section 5.

\section{Conditional Probability Interpretation}

In the Conditional Probability Interpretation (CPI) of quantum mechanics, there is no such thing as a direct measurement of time.
The notion of measuring time is expressed through the use of a physical clock. A clock is simply a physical system, and its
variables (position, momentum...) are what we measure, and use as references, or ``time''. We usually choose the clock to be the
least correlated with the system under study, so that a measurement of the clock variables will not greatly affect a measurement
in the system of interest. Being a physical system, a clock can be fully described by a Hamiltonian, and since, from now on,
we will assume the clock is fully uncorrelated with the system (which practically can never be exactly achieved), we can write
\begin{equation}
\label{Hamiltonian }
H_{\rm tot}=H_{\rm Clock}+H_{\rm System}.
\end{equation}
The action then becomes 
\begin{equation}
\label{Action }
S_{\rm tot}=\int \textrm{d}n \textrm{ } \left( L_{\rm Clock}(n)+L_{\rm System}(n)\right).
\end{equation}

In the above equation, $n$ is a parameter, not a variable, and as such it can not be measured. In a certain way, it could be seen as
some abstract internal time. This parameter ensures the unitarity of the quantum theory emerging from the CPI. It may not be unitary with
respect to the time measured by a physical clock, but it will stay unitary with respect to this ``internal time''. This is important since
it indicates that the CPI does not in fact question one of the fundamental pillar of Quantum Mechanics.  Also $n$ defines simultaneity.
Two events are said to be simultaneous if they happen for the same value $n$. 
In the CPI, the probability of measuring a system in a state $|o\rangle$ at a time $t$, becomes the probability of measuring the system
in a state $|o\rangle$  \textit{when} measuring the clock in a state $|t\rangle$, and for a closed system is expressed as \cite{relational_solution}
\begin{equation}
\label{single time probability }
\mathcal{P}(o\in \Delta o | t\in \Delta t)=\frac{\int_{n}\langle\psi|P_{t}(n)P_{o}(n)P_{t}(n)|\psi\rangle}{\int_{n}\langle\psi|P_{t}(n)|\psi\rangle},
\end{equation} 
where $|\psi\rangle$ is the initial state (at some $n_{o}$) of the total system (\textit{clock-system of interest}). The projectors are defined by
\begin{eqnarray}
P_{o}(n)=\int_{\Delta o}\int_{i}|o,i,n\rangle\langle o,i,n| \nonumber \\
P_{t}(n)=\int_{\Delta t}\int_{j}|t,j,n\rangle\langle t,j,n| \nonumber.
\end{eqnarray}

We have assumed in the previous set of equations that the eigenvalues of the operators $\widehat{O}$ and $\widehat{T}$ have continuous spectra,
which usually will be the case. However for now on, we will drop the integral over the interval $\Delta o$ and $\Delta t$, in an attempt to
simplify the notation. That is to say we consider the spectrum to be discrete. The future calculations will not be greatly affected by this
approximation. Since the intervals in questions are very small, the results will only differ by factors of $\Delta o$ and $\Delta t$.
Those factors will be absorbed in the normalization of the probability. So as long as we keep in mind that $\int\textrm{d}o\mathcal{P}(o|t)=1$,
we can forget that the spectra are continuous.  Also, in order to keep the equation more concise we assume the eigenvalues of
$\widehat{O}$ and $\widehat{T}$ form a complete set. We can now rewrite the projectors as
\begin{eqnarray}
P_{o}(n)=|o,n\rangle\langle o,n| \nonumber \\
P_{t}(n)=|t,n\rangle\langle t,n| .
\end{eqnarray}
Since the clock and the system under study are assumed to be fully uncorrelated, the operators associated with the clock commute
with the operators associated with the system. Therefore we can split a vector into a\textit{ clock vector }and a \textit{system vector},
\begin{equation}
\label{vector separation}
|\psi\rangle=|\psi_{c}\rangle\otimes|\psi_{s}\rangle .
\end{equation}
In the same fashion,
\begin{equation}
\label{projection operator fusion}
P_{t}(n)P_{o}(n)P_{t}(n)=P_{t,o}(n)=(|t,n\rangle\otimes|o,n\rangle)(\langle t,n|\otimes\langle o,n|).
\end{equation}
The parameter $n$ replaces $t$ as the parameter of the action, so as in ``Standard Quantum Mechanics'' (SQM) we can define a unitary
operator $\widehat{U}(n_{f}-n_{i})=e^{-\frac{i}{\hbar}H(n_{f}-n_{i})}$ which will evolve operators from $n_{i}$ to $n_{f}$:
\begin{equation}
\label{evolution operator}
P_{t,o}(n)=\widehat{U}^{\dag}(n)P_{t,o}\widehat{U}(n).
\end{equation}

With this framework now defined, we can make more meaningful calculations, for example the probability corresponding to a
two-time measurement, which is simply the square of the propagator in SQM. This particular calculation was used by Kuchar to argue
against the CPI~\cite{kuchar}. However in Kuchar's approach of the CPI, the parameter $n$ was missing, forbidding the system to evolve.
The correct expression for the probability of measuring $o$ at $t$ when we have $o'$ at $t'$, is given in Ref.~\cite{relational_solution}:
\begin{equation}
\label{two time probability }
\mathcal{P}(o|to't')=\frac{\int_{nn'}\langle\psi_{c}|P_{t'}(n')P_{t}(n)P_{t'}(n')|\psi_{c}\rangle\langle\psi_{s}|P_{o'}(n')P_{o}(n)P_{o'}(n')|\psi_{s}\rangle}{\int_{nn'}\langle\psi_{c}|P_{t'}(n')P_{t}(n)P_{t'}(n')|\psi_{c}\rangle\langle\psi_{s}|P_{o'}(n')|\psi_{s}\rangle}.
\end{equation}
Using (\ref{projection operator fusion}), (\ref{evolution operator}) and defining $U_{s}(n)=e^{-\frac{i}{\hbar}H_{s}(n)}$, $U_{c}(n)=e^{-\frac{i}{\hbar}H_{c}(n)}$, we find
\begin{equation}
\label{num_o}
\langle\psi_{s}|P_{o'}(n')P_{o}(n)P_{o'}(n')|\psi_{s}\rangle = |\psi_{s}(o',n')|^{2}|\langle o'|U_{s}(n'-n)|o\rangle|^{2},
\end{equation}
where $\psi_{s}(o',n')=\langle o'|U_{s}(n')|\psi_{s}\rangle$. Similarly,
\begin{equation}
\label{num t }
\langle\psi_{c}|P_{t'}(n')P_{t}(n)P_{t'}(n')|\psi_{c}\rangle = |\psi_{c}(t',n')|^{2}|\langle t'|U_{c}(n'-n)|t\rangle|^{2},
\end{equation}
and 
\begin{equation}
\label{ den o}
\langle\psi_{s}|P_{o'}(n')|\psi_{s}\rangle = |\psi_{s}(o',n')|^{2}.
\end{equation}
We can now write the final expression for the conditional probability,
\begin{equation}
\label{conditional probability }
\mathcal{P}(o|to't')=\frac{\int_{nn'}|\psi_{s}(o',n')\psi_{c}(t',n')\langle t'|\widehat{U}_{c}^{\dag}(n-n')|t\rangle\langle o'|\widehat{U}_{s}^{\dag}(n-n')|o\rangle|^{2}}{\int_{nn'}|\psi_{s}(o',n')\psi_{c}(t',n')\langle t'|\widehat{U}_{c}^{\dag}(n-n')|t\rangle|^{2}},
\end{equation}
and one can easily check
\begin{equation}
\label{normalization }
\int \textrm{d}o \mathcal{P}(o|to't')=1
\end{equation}

In SQM, the probability of measuring a variable $o$ at time $t$ when we measured $o'$ at time $t'$ is given by the
propagator $K(ot,o't')=\langle o'|\widehat{U}_{s}^{\dag}(t-t')|o\rangle$. In the CPI the propagator is replaced by the
conditional probability. At first sight it is not obvious that the two descriptions are equivalent. However, the
success of SQM demands that the CPI must give a propagator that recovers the SQM propagator in the
limit of today's experimental accuracy.

\section{Semi-Classical Clock}

To recover SQM from the CPI, one has to choose a clock that behaves almost classically. Two things will dictate such behavior. First, the
clock itself. Some clocks will allow a more classical regime than others. Second is the clock's initial state (at a given $n_{o}$). Indeed,
this can be easily understood from the uncertainty principle. When we measure a variable very precisely, the uncertainty in its canonical
conjugate will be great. Since the evolution of a variable usually depends on its canonical conjugate (thought not always as we will see),
the measurement after a certain ``time'' (internal time $n$) will be meaningless.
For example, if we use a free particle as a clock, associating time with its position, and we start with a definite value for its
initial position, 
\begin{equation}
\label{initial position state}
|\psi_{s}\rangle=\int\textrm{d}x \, \delta(x-x_{0})|x\rangle,
\end{equation}
the wavefunction in momentum space will obviously indicate that there is equal probability for the momentum to be taking any value
(of course we are not treating the particle as being relativistic since Quantum Field Theory would need to be used in such cases).
At the second measurement (that is at a greater $n$), we have equal probability of finding the particle at any position. That makes for a
very poor clock, and definitely not a classical clock.

In Quantum Mechanics, generally the recovery of classical results is not straightforward. For example, the free particle propagator is
usually not the same as what is predicted by Classical Mechanics:
\begin{equation}
\label{classical propagator}
K(xt,x_{o}t_{o})=\delta\big(x-(x_{o}+\frac{p^{2}}{m})\big).
\end{equation}

However it is possible to find a propagator in the Quantum regime that mimics the classical one closely by using coherent states. 
Instead of a delta function, we choose the state of the particle in momentum space, at time $t_{o}$, to be a Gaussian distribution of
width $\sigma_{p}$ peaked around a value $p_{o}$,
\begin{equation}
\label{gaussian momentum distribution}
|\psi\rangle \propto \int \textrm{d}p \, e^{-\frac{i}{\hbar}\big(\frac{p-p_{o}}{2\sigma_{p}}\big)^{2}}|p\rangle.
\end{equation} 
In the position basis, we also get a Gaussian distribution of width $\sigma_{x}=\frac{\hbar}{2 \sigma_{p}}$, peaked around 0.
We evolve the system through a time $t$ to find
\begin{equation}
\label{gaussian momentum evolved}
|\psi(t)\rangle \propto \int \textrm{d}p \, e^{-\frac{i}{\hbar}\frac{p^{2}}{2m}t}e^{-\frac{i}{\hbar}\big(\frac{p-p_{o}}{2\sigma_{p}}\big)^{2}}|p\rangle.
\end{equation}
We finally express the wavefunction in the position basis by Fourier transforming,
\begin{equation}
\label{evolved postion distribution}
|\psi(t)\rangle \propto \int \textrm{d}x \, e^{-\frac{i}{\hbar}\big(\frac{x-\frac{p_{o}t}{m}}{2\sigma(t)}\big)^{2}}|x\rangle.
\end{equation}
The Gaussian in the position domain is peaked around $p_{0} t/ m$, the classical distance traveled by the particle. 

Also, we know the non-classicality of the clock comes from the uncertainty principle between the variable we use as a measure of time, and
its canonical momentum. Coherent states minimize the uncertainty: $\sigma_{x}\sigma_{p}=\frac{\hbar}{2}$. Therefore we expect the clock to
be the most classical when using a coherent state.
To illustrate the dependence of the classical regime to the type of clock and to the initial state of the clock, we will go through two examples,
the parameterized Hamiltonian and the free particle.

\subsection{Parameterized Hamiltonian}

This particular clock has already partly been studied by Dolby, who used a different formalism of the CPI~\cite{dolby}. The clock's Hamiltonian is linear in its
generalized momentum $k$. The system under study is described by a general Hamiltonian $H_{s}$:
\begin{equation}
\label{parametrized hamiltonian}
H=k+H_{s}.
\end{equation}
Let's solve the classical equation of motion for the clock to have an idea of what to expect. The action is defined as
\begin{equation}
\label{parametrized action}
S=\int\textrm{d}n\big[k \dot{t}-k+L_{s}(q,\dot{q})\big],
\end{equation}
where $\dot{t}=\frac{\textrm{d}t}{\textrm{d}n}$. The equations of motion are simply $\dot{p}=0$ and $\dot{t}=1$. Then, up to a constant, $t=n$.
We see that the variable $t$ we choose to measure is exactly equal to the internal time $n$, and its evolution is not dictated by its momentum.
We can already guess that in the Quantum regime, the uncertainty principle won't affect the evolution of the operator $\widehat{T}$ associated with
the variable $t$. Then if we start with a state $|\psi_{t}\rangle=|t=0\rangle$, the delta function will not spread and will remain sharp, the clock
therefore remains classical. From (\ref{conditional probability }), we have
\begin{equation}
\label{two time probability 2}
\mathcal{P}(o|to't')=\frac{\int_{nn'}|\psi_{s}(o',n')\psi_{c}(t',n')\langle t'|\widehat{U}_{c}^{\dag}(n-n')|t\rangle\langle o'|\widehat{U}_{s}^{\dag}(n-n')|o\rangle|^{2}}{\int_{nn'}|\psi_{s}(o',n')\psi_{c}(t',n')\langle t'|\widehat{U}_{c}^{\dag}(n-n')|t\rangle|^{2}}.
\end{equation}  
Replacing the clock Hamiltonian by $H_{c}=k$, and using the completeness of the basis, we get
\begin{equation}
\label{propagator part}
\langle t'|e^{-\frac{i}{\hbar}k(n'-n)}|t\rangle=\int\textrm{d}k \langle t'|k\rangle e^{-\frac{i}{\hbar}k(n'-n)}\langle k|t'\rangle=\delta\big((n'-n)-(t'-t)\big).
\end{equation}
Using (\ref{two time probability 2}), we finally find
\begin{equation}
\label{quantum classical propagator}
P(o|t,o',t')=\big|\langle o'|U_{s}(t'-t)|o\rangle\big|^{2}=\big|K(ot,o't')\big|^{2},
\end{equation}
as expected. We see that the propagator from SQM is recovered from the CPI in the case of the parameterized Hamiltonian for any initial state of the clock.
However the recovery of the whole SQM without any dependence on the shape of the initial state is not possible. Indeed the one time probability in the
CPI is given by (\ref{single time probability }):
\begin{equation}
\label{single time probability 2}
\mathcal{P}(o|t)=\frac{\int\textrm{d}n\big|\langle\psi_{c}|U_{c}^{\dagger}(n)|t\rangle\big|^{2}\big|\langle\psi_{s}|U_{s}^{\dagger}(n)|o\rangle\big|^{2}}{\int\textrm{d}n\big|\langle\psi_{c}|U_{c}^{\dagger}(n)|t\rangle\big|^{2}}.
\end{equation}
Or, in the same notation as Ref.~\cite{fundamental_decoherence}:
\begin{equation}
\label{single time probability 3}
\mathcal{P}(o|t)=\int_{n}\big|\langle\psi_{s}|U_{s}^{\dagger}(n)|o\rangle\big|^{2}\mathcal{P}_{n}(t)=\int\textrm{d}n\big|\langle\psi_{s}|U_{s}^{\dagger}(n)|o\rangle\big|^{2}\frac{\big|\int\textrm{d}k e^{\frac{i}{\hbar}k(n-t)}\langle\psi_{c}|k\rangle\big|^{2}}{\int\textrm{d}n\big|\int\textrm{d}k e^{\frac{i}{\hbar}k(n-t)}\langle\psi_{c}|k\rangle\big|^{2}},
\end{equation}
where $\mathcal{P}_{n}(t)$ is the probability that the ``internal time'' takes the value $n$ when we measure $t$. The SQM equivalence,
$\mathcal{P}(o|t)=\big|\langle\psi_{s}|U_{s}^{\dagger}(t)|o\rangle\big|^{2}$, is only recovered for
$\mathcal{P}_{n}(t)=\delta(n-t)$, which is achieved for
\begin{equation}
\label{ideal clock limit }
|\psi_{c}\rangle=\int dt \, \delta(t)|t\rangle.
\end{equation}
This agrees with the result found by Dolby, who calls $\psi_{c}\rightarrow\delta(t)$ the ``ideal clock limit''. It is certainly true for this specific
type of clock, but it won't necessarily be an ideal limit for every clock, as we shall see. In this light, SQM is a non-physical limit to the CPI. It implies the use of a non-physical clock.

\subsection{Free Particle}

Here we use the position of a free particle of mass $m$ as a measure of time. The Hamiltonian for the clock is 
\begin{equation}
\label{free particle hamiltonian}
H_{c}=\frac{p^{2}}{2m}.
\end{equation}
This particular clock was studied in a recent paper by Gambini {\it et al.}~\cite{Conditional_probabilities_dirac}. There, the authors approach the problem in a slightly different way. They consider the Hamiltonian to be parametrized, their goal being to show that the CPI behaves well for a constraint system. They also mention that SQM can be recovered only to leading order. Here, we will not worry about constraints but rather focus on how closely the SQM can be recovered. We will then use this result in section 4 to calculate the minimum decoherence one could achieve with this simple clock.

If we calculate the single time probability $\mathcal{P}(o|x)$ using Dolby's ``ideal clock limit'', we find 
\begin{equation}
\label{ideal free particle part}
\langle\psi_{c}|U_{c}^{\dag}(n)|x\rangle=\sqrt{\frac{2m\hbar\pi}{i}}e^{\frac{it^{2}m}{2\hbar n}},
\end{equation}
and then the probability becomes
\begin{equation}
\label{ ideal free particle }
\mathcal{P}(o|x)=\frac{1}{V_{n}}\int_{n}\big|\langle\psi_{s}|U_{s}^{\dag}(n)|o\rangle\big|^{2}
\end{equation}
If we were to measure $x$ for the particle position, there would be an equal probability that it corresponds to any possible value for
the internal time. Therefore the time measured does not give any indication on the internal time. This is {\em not} an ideal clock. In
fact $\mathcal{P}_{n}(t)=\delta(t-n)$ (with $t=m x /p_{o}$) cannot be achieved for such a clock, since any wave packet in position space
will spread due to the position-momentum uncertainty. However it is possible to get a probability distribution sufficiently peaked to approximate
a delta function. In order to minimize the peak's width in both position and momentum, we use a coherent state:
\begin{equation}
\label{coherent state}
|\psi_{c}\rangle=\int\textrm{d}p \frac{1}{(2\pi)^{1/4}\sqrt{\sigma_{p}}}e^{-\big(\frac{p}{2\sigma_{p}}\big)^{2}}|p+p_{o}\rangle=\int\textrm{d}x \frac{1}{(2\pi)^{1/4}\sqrt{\sigma_{x}}}e^{-\big(\frac{x}{2\sigma_{x}}\big)^{2}}|x\rangle,
\end{equation} 
where $\sigma_{x}\sigma_{p}=\frac{\hbar}{2}$, and where we centered the Gaussian in momentum space around the classical momentum $p_{o}$.
Using these expressions, and performing some algebra, we find the probability distribution to be a Gaussian  
\begin{equation}
\label{free particle distribution}
\mathcal{P}_{n}(x)\alpha e^{-\frac{1}{2}\big(\frac{x-\frac{p_{o}}{m}n}{\delta(n)}\big)^{2}},
\end{equation}   
of width $\delta(n)=\sqrt{\sigma_{x}^{2}+\frac{\sigma_{p}^{2}n^{2}}{m^{2}}}$. The width can be minimize with respect to $\sigma_{x}$,
taking into account that the width of the Gaussians in position and momentum are related through the uncertainty principle by
$\sigma_{x}\sigma_{p}=\frac{\hbar}{2}$. The minimum will occur at $\sigma_{x}^{2}=\frac{\hbar n}{2m}$ and will take the value
\begin{equation}
\label{distribution minimum width}
\delta_{min}(n)=\sqrt{\frac{\hbar n}{m}}.
\end{equation}
The optimum initial state for the clock will depend on ``when'' (at which internal time) we are making the measurement. Of course this presents
a problem since one cannot tell ``when'' the measurement on the clock is taken before taking it, and even then one will have only a peaked
probability distribution as an indication of what the final internal time is. It is therefore not possible to ``prepare'' the clock in order to
ensure a minimal spread for $\mathcal{P}_{n}(x)$. To conclude this example, we note that for a free particle clock, SQM is only recovered
on scales larger than $\frac{\hbar n}{m}$, and even then the recovery will only be partial, since some small deviation from the SQM will
still occur as we will see in the next section.

\section{Limitation in the accuracy of a clock and Decoherence}

The CPI implies a non-unitarity of the theory with respect to the ``time'' measured through the physical clock. This in turn implies that the
system under study will not evolve as SQM predicts. Rather, a Lindblad type equation~\cite{lindblad} describes its evolution~\cite{relational_solution,isidro},
\begin{equation}
\label{Lindblad equation}
\frac{\partial\tilde{\rho}_{s}(t)}{\partial t}=-\frac{i}{\hbar}\big[(1+\beta(t))\widehat{H}_{s},\tilde{\rho}_{s}\big]-\sigma(t)\big[\widehat{H}_{s},[\widehat{H}_{s},\tilde{\rho}_{s}]\big].
\end{equation}
Here $\tilde{\rho}_{s}$ is the corrected density matrix of the system under study. It is corrected in the sense that it satisfies the usual
equation for a single time probability,
\begin{equation}
\label{tilde single time probability}
\mathcal{P}(o|t)=\frac{Tr\big(P(o)\tilde{\rho}(t)\big)}{Tr\big(\tilde{\rho}(t)\big)}.
\end{equation} 
Instead of modifying this equation as we did earlier, we have defined a new density operator $\tilde{\rho}_{s}$. It is the density matrix in the CPI regime.

The second term in (\ref{Lindblad equation}) is the major point of departure from the standard Heisenberg evolution equation for Quantum operators.
Due to this term, a system will loose information upon evolution. The system is said to decohere. The decoherence factor $\sigma(t)$ is closely
related to the probability distribution $\mathcal{P}_{n}(t)$. However Gambini and Pullin showed it only depends on the spread $b(t)$ and the asymmetry
of the distribution~\cite{real_rods}. If we assume there is no asymmetry, which is true for the clocks we studied, then the decoherence factor is given by 
\begin{equation}
\label{decoherence factor}
\sigma(t)=\frac{\partial}{\partial t}b(t)
\end{equation}
 
 In order to give an estimate for the fundamental decoherence, Gambini and Pullin used a limit on the accuracy of physical clocks found by Ng and van Dam.
This limit was found using a simple clock composed of two mirrors and a photon bouncing between them. Using SQM and the uncertainty in the position of the
mirror, they argued that the time it takes for the photon to travel from one mirror to the other can not be measured exactly. This in turn implies
that there is a limit in the accuracy of spatial measurement, given by~\cite{limitation}
 \begin{equation}
\label{Ng limitation}
\delta x=\delta x(0)+\delta x(n)= \delta x(0) +\frac{1}{2}\frac{\hbar t}{m\delta x(0)}.
\end{equation}
This was used to calculate a minimum decoherence (after minimization with respect to $\delta x(0)$). However we believe that the CPI is self-consistent, and
the minimum spread in a clock accuracy (and consequently a minimum decoherence) can be found without relying on any SQM result.

First we note that the clock used by Ng and van Dam is really measuring a distance, since they assume there is no uncertainty in the time measured for a
photon to cover a given distance. Therefore their time $t$ is equivalent to our internal time $n$. Since the clock gives us the spatial separation between
the two mirror, it is equivalent to a free particle in the CPI, the particle taking the role of the mirror. There is only one difference between the two
pictures. In the free particle clock, the mirror's momentum is peaked around a classical value $p_{o}$ instead of being classically stationary. But since
being stationary correspond to the special case $p_{o}=0$, the precision of the measurement will be unchanged.
Then we realize that the particle wavefunction in the position representation depends only on the probability distribution $\mathcal{P}_{n}(x)$,
\begin{equation}
\label{position wrt distribution}
|\psi(n)\rangle=e^{-\frac{ip_{o}^{2}n}{2m\hbar}}\int\textrm{d}x\sqrt{\mathcal{P}_{n}(x)}|x\rangle.
\end{equation}
 We already calculated $\mathcal{P}_{n}(x)$ for a free particle (\ref{free particle distribution}),
\begin{equation}
\label{free particle distribution 2}
\mathcal{P}_{n}(x)\alpha e^{-\frac{1}{2}\big(\frac{x-\frac{p_{o}}{m}n}{\delta_{x}(n)}\big)^{2}},
\end{equation}
where $\delta_{x}$ is
\begin{equation}
\label{delta x}
\delta_{x}=\big|\langle x^{2}\rangle-\langle x\rangle^{2} \big|^{1/2}=\sqrt{\delta^{2}_{x}(0)+\frac{\hbar^{2}n^{2}}{4\delta^{2}_{x}(0)m^{2}}}.
\end{equation}
 
We immediately see that our result differs from Ng and van Dam's. Instead of adding the uncertainty at $n_{\rm initial}$ with the one at $n_{\rm final}$,
we sum their squares. This discrepancy is conceptually important. However upon minimization of $\delta_{x}$ with respect to $\delta_{x}(0)$, the
two versions agree,
 \begin{equation}
\label{minimum spread}
\delta x_{min}=\sqrt{\frac{\hbar n}{m}}.
\end{equation}  
One can find the decoherence's strength associated, using $b(t)=(\delta x_{min})^{2}_{n=t}$,
\begin{equation}
\label{minimum decoherence}
\sigma(t)=\frac{\hbar}{m}.
\end{equation}

The CPI enables us to recover an important result, in a very fundamental way. We started from a simple Hamiltonian, and established without any
assumption and without using any results from SQM that there is indeed a limit to the clock accuracy, which in turn will induce a loss of information
in any system studied, through decoherence. This is a very powerful method since we can now imagine doing a similar calculation with a more realistic
clock (pendulum or particle decay). All we need is the Hamiltonian for such a clock.

One last point worth noting is the lack of decoherence when using the (un-realistic) clock described by the Hamiltonian $H_{c}=p$. Indeed the
spread $b(t)$ in the distribution $\mathcal{P}_{n}(t)$ does not depend on $t$. It is not surprising since this clock is the closest to a classical clock.
This lack of decoherence is exact for the ``ideal clock limit'' (\ref{ideal clock limit }), but is also observed for a general initial state as long
as we consider sufficiently large time. By large we mean much larger than the spread of $\mathcal{P}_{n}(t)$. If we are interested in time scales similar
to $\sqrt{b(t)}$, then we cannot approximate our distribution with delta functions:
\begin{equation}
\label{delta function approximation}
\mathcal{P}_{n}(t)\neq\delta(n-t)+b(t)\delta''(n-t).
\end{equation}
One has to use the full expression for the probability distribution,
\begin{equation}
\label{full probability distribution}
\mathcal{P}_{n}(t)=\frac{\sqrt{2}e^{-\frac{(t-n)^{2}}{2\sigma_{t}^{2}}}}{\sqrt{\pi}\sigma_{t}{\rm erfc}(-\frac{t}{\sqrt{2}\sigma_{t}})}.
\end{equation}
Here the initial state was a Gaussian of width $\sigma_{t}$ in time. To calculate the magnitude of the decoherence,
we use a density matrix expressed in its eigenenergy basis:
\begin{equation}
\label{rhotilde}
\tilde{\rho}_{s}(t)=\int_{0}^{\infty} \textrm{d}n e^{-i\frac{\hat{H}_{s}n}{\hbar}} \rho_{s} e^{i\frac{\hat{H}_{s}n}{\hbar}}\mathcal{P}_{n}(t),
\end{equation}
and 
 \begin{equation}
\label{densityeigenenergy}
\rho_{s}=\int\int\textrm{d}E\textrm{d}E' A_{EE'}|E\rangle\langle E'|.
\end{equation}
This choice allows us to explicitly perform the integration over the parameter $n$, which greatly simplifies the calculation. We find
 \begin{equation}
\label{rhotilde2}
\tilde{\rho}_{s}(t)=e^{-i\frac{\hat{H}_{s}t}{\hbar}} \tilde{\rho}_{o} e^{i\frac{\hat{H}_{s}t}{\hbar}},
\end{equation} 
with 
\begin{equation}
\label{rhoo}
\tilde{\rho}_{o}(t)=\int\int\textrm{d}E\textrm{d}E' \frac{A_{EE'}|E\rangle\langle E'|}{erfc\big(\frac{-t}{2\sigma_{t}}\big)}erfc\Big(\frac{E-E'}{2\sigma_{E}}i-\frac{t}{2\sigma_{t}}\Big)e^{-\big(\frac{E-E'}{2\sigma_{E}}\big)^{2}}.
\end{equation}
Consequently, the evolution of the density operator is given by
\begin{equation}
\label{evolutionpluscorrection}
\frac{\partial\tilde{\rho}_{s}(t)}{\partial t}=\frac{i}{\hbar}[\tilde{\rho}_{s}(t),\widehat{H}_{s}]+e^{-i\frac{\hat{H}_{s}t}{\hbar}} \frac{\partial\tilde{\rho}_{o}(t)}{\partial t} e^{i\frac{\hat{H}_{s}t}{\hbar}},
\end{equation}
 which is the Heisenberg equation of motion plus a correction. We are interested in the magnitude of this correction, and especially in the rate at
which it dies off as $t$ becomes large compared to $\sigma_{t}$. We explicitly calculate $\frac{\partial\tilde{\rho}_{o}(t)}{\partial t}$:
\begin{equation}
\label{rhooevolution}
\frac{\partial\tilde{\rho}_{o}(t)}{\partial t}=\int\int\textrm{d}E\textrm{d}E' A_{EE'}|E\rangle\langle E'|f(t),
\end{equation}
with
\begin{equation}
\label{correctionmagnitude}
f(t)=\frac{e^{-\big(\frac{t}{2\sigma_{t}}\big)^{2}-\big(\frac{E-E'}{2\sigma_{E}}\big)^{2}}}{\sqrt{\pi}\sigma_{t}erfc\big(\frac{-t}{2\sigma_{t}}\big)}\left[ \frac{ e^{-\big(\frac{t}{2\sigma_{t}}-\frac{E-E'}{2\sigma_{E}}i\big)^{2}}}{e^{-\big(\frac{t}{2\sigma_{t}}\big)^{2}}} - \frac{erfc\big(\frac{E-E'}{2\sigma_{E}}i-\frac{t}{2\sigma_{t}}\big)}{erfc\big(-\frac{t}{2\sigma_{t}}\big)} \right] .
\end{equation}
 
We first notice that the magnitude of the correction to the evolution equation depends on the density matrix element we are looking at. In particular,
the correction vanishes for the diagonal terms. In general, the correction factor will stay small for nearly diagonal elements ($\|E-E'\|<\sigma_{E}$),
and will be constant throughout the rest of the matrix. Also $f(t)$ vanishes when $\sigma_{t}=0$ for any nonzero $t$. This was expected since we fully
recover SQM in that limit. The correction function will vanish as well for values of $t$ much bigger than the spread $\sigma_{t}$ in the clock's initial
state ($t>3\sigma_{t}$). The system will therefore decohere for an amount of time that depends on the spread of the probability distribution
$\mathcal{P}_{n}(t)$, which for this Hamiltonian is equivalent to how well localized is the clock's initial state in the $t$ representation.
For times larger than $t \sim 3\sigma_{t}$ the system will undergo a standard evolution. This type of decoherence will be present for any type of clock,
but will die off in a similar fashion and can be neglected for large enough time scales.
 
 \section{Conclusion}
 In this paper we have shown that a classical clock can be described by an Hamiltonian linear in momentum. Even though the initial state of the clock must
be a delta function in the time variable space (Dolby's ``ideal clock limit'') in order for the clock to be fully classical, the dynamical features of
SQM (two-time probability or propagator) is recovered no matter what initial state is used, which negates Kuchar's objection to the CPI.
This is also the case for the lack of decoherence at sufficiently large scale. However, even for the simplest physically realistic clock, a free particle,
the classical limit can not be recovered, and the use of the ``ideal clock limit'' will actually move the clock away from its classical regime. We showed
that even though this clock cannot behave classically, there exist a semi-classical regime in which the CPI discrepancies with SQM are kept to a minimum.
It can be achieved by using an initial state that minimizes the uncertainty relation between the ``time variable'' and its canonical momentum. This state is
said to be coherent, and its distribution in the variable or its associated momentum space is a Gaussian. We then used those coherent states to calculate the
minimum decoherence one can achieve for a ``free particle clock'', and we found our results to be in agreement with previous estimates~\cite{limitation}.
We also found out that one cannot be sure the decoherence one observes is minimal. Indeed the initial state needed to get closest to the classical regime
depends on the internal time $n_{\rm final}$ at which the final measurement is taken. Since $n_{\rm final}$ cannot be exactly predicted, one cannot
``prepare'' the clock in advance in order to achieve the minimal decoherence.
 
Similar calculation could be carried for more realistic clocks, such as a pendulum, or a particle decay, in order to fully understand the implications of
the CPI in concrete experiments. This is particularly important in the view of the development of Quantum information theory, and Quantum computing.
To do so, one would need to generalize the above calculation to include the use of ``rods'', in addition to ``clocks'', to measure distances and therefore
be able to study systems traditionally described by Quantum Fields. The use of measuring ``rods'' in Quantum Field Theory has been discussed in
Ref.~\cite{rods}.  The effect of decoherence in dramatically curved space would also be essential in the understanding of black holes and of the early
stages of our Universe.

\bibliography{decoherencebib}	
 
\end{document}